\documentclass[12pt,english]{elsarticle}
\usepackage[LGR,T1]{fontenc}
\usepackage[latin9]{inputenc}
\usepackage{geometry}
\usepackage{float}
\usepackage{units}
\usepackage{textcomp}
\usepackage{amsmath}
\usepackage{graphicx}

\makeatletter

\DeclareRobustCommand{\greektext}{%
  \fontencoding{LGR}\selectfont\def\encodingdefault{LGR}}
\DeclareRobustCommand{\textgreek}[1]{\leavevmode{\greektext #1}}
\ProvideTextCommand{\~}{LGR}[1]{\char126#1}

\newcommand{\lyxmathsym}[1]{\ifmmode\begingroup\def\b@ld{bold}
  \text{\ifx\math@version\b@ld\bfseries\fi#1}\endgroup\else#1\fi}

\providecommand{\tabularnewline}{\\}

\date{}

\makeatother

\usepackage{babel}
\begin{document}

\begin{frontmatter}{}

\title{Complex dielectric behaviours in $BiFeO_{3}/Bi_{2}Fe_{4}O_{9}$ ceramics}

\author[1]{Gilad Orr}
\ead{gilad.orr@ariel.ac.il}
\author[1]{Andrey Gorychev} 
\author[1]{Paul Ben Ishai\corref{cor1}}
\ead{paulbi@ariel.ac.il}
\cortext[cor1]{Corresponding author paulbi@ariel.ac.il}
\address[1]{Department of Physics, Ariel University, Ariel 40700, Israel}
\begin{abstract}
The complex dielectric permittivity of a sintered ceramic tablet consisting
of 70.5 \% $BiFeO_{3}$ and 27.7\% $Bi_{2}Fe_{4}O_{9}$ was analyzed
as a function of temperature from -120 °C to 230 °C. The results reveal
a complicated dielectric response with temperature activated relaxation
features. They also reveal a ferroelectric phase transition that decayed
with repeated heating cycles of the tablet. The source of the behaviour
is assigned to relaxation processes happening along the grain boundaries
of differing compositions in the tablet. The origin of the phase transition
is traced to locally induced strains on grain boundaries because of
unit cell size mismatch between $BiFeO_{3}$ and $Bi_{2}Fe_{4}O_{9}$.
\end{abstract}

\end{frontmatter}{}

\section{Introduction}

\begin{figure}[h]
\begin{centering}
\includegraphics[scale=0.5]{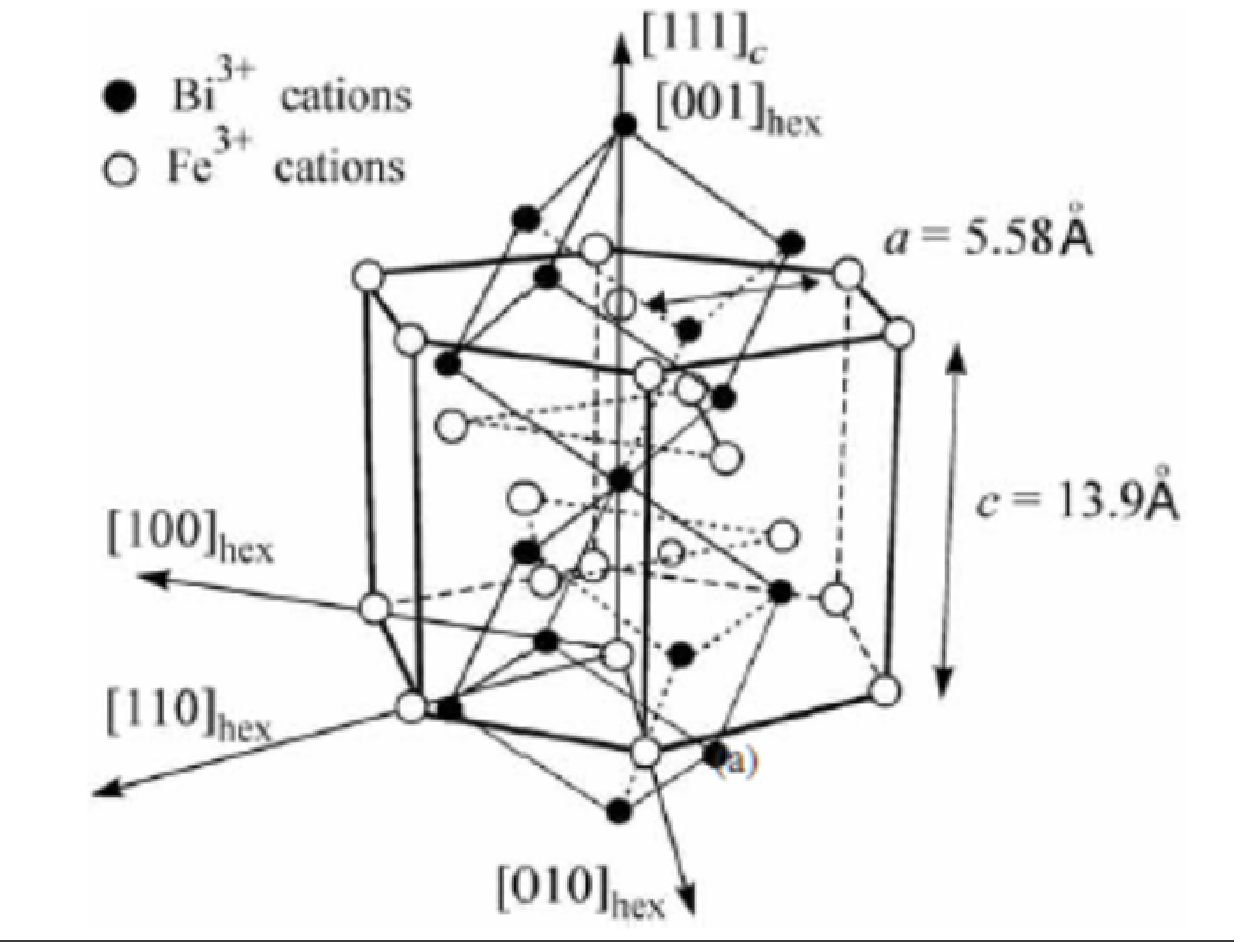}
\par\end{centering}
\caption{\label{fig:The-unit-cell}The unit cell of BiFeO3 (reproduced from
Ref. \citep{kadomtseva2006phase} ) the axis directions are given
for both the cubic and hexagonal frames of reference. For clarity
the oxygen ions are not shown.}
\end{figure}

Bismuth Ferrous oxides are the poster boys for multi-ferroics, a class
of materials capable of simultaneously hosting both ferromagnetic
and ferroelectric properties and first predicted by P. Curie \citep{curie1894symetrie}.
Of these Bismuth Ferrite, BiFeO3 , first studied seriously by Smolensky
in 1958 and later produced as a single phase \citep{achenbach1967preparation},
has taken the podium position as the archetypal multi-ferroic \citep{velasco2016synthesis},
\citep{catalan2009physics}. The unit cell of Bismuth Ferrite is a
perovskite type structure at room temperature belonging to the space
group R3c \citep{michel1969atomic}, \citep{karpinsky2017thermodynamic}
with 2 formula units involved in the basic unit cell \citep{kadomtseva2006phase},
reproduced in Figure \ref{fig:The-unit-cell}, resulting in a rhombohedral.

\begin{figure}[H]
\begin{centering}
\includegraphics[scale=0.7]{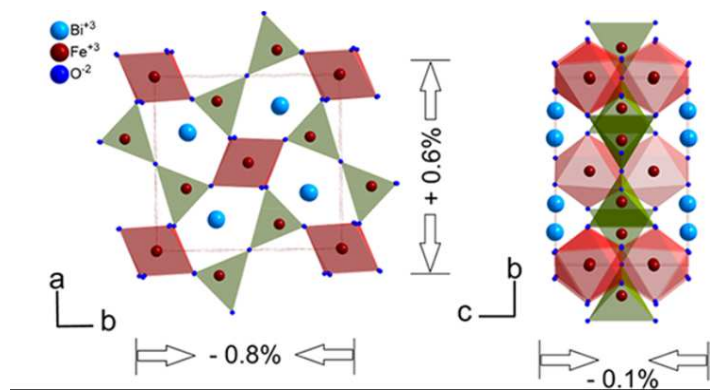}
\par\end{centering}
\caption{\label{fig:A-comparison-of}A comparison of the unit cell sizes between
$BFO$ and mullite-type $Bi_{2}Fe_{4}O_{9}$. reproduced from \citep{kirsch2019structural}}
\end{figure}

For clarity the oxygen atoms are not shown. They occupy the face centers
of the Bismuth cubes. As has been noted this structure results in
a room temperature ferroelectric polarization along the {[}111{]}cubic
axis as well as a distortion to the oxygen octahedra by rotating it
approximately 11-14° around the {[}111{]} axis \citep{catalan2009physics}.
This leads to an Fe-O-Fe angle of $\theta=154\lyxmathsym{\textendash}156\text{\textdegree}$
\citep{catalan2009physics}, the significance of which is to control
both the magnetic ordering and orbital overlap between Fe and O. This
is the source of BFO\textquoteright s magnetic behavior. The ferroelectric
transition \citep{catalan2009physics} is at T=825 °C, meaning that
the pure BFO is ferroelectric and anti-ferromagnetic (type G) at room
temperature \citep{karpinsky2017thermodynamic}, \citep{wang2020structure}.
Unlike most inorganic perovskites the polarization axis is along the
diagonal {[}111{]} perovskite direction \citep{shvartsman2007large},
while its magnetic axis is perpendicular to the same axis. Indeed,
its intrinsic polarization along this access has been measured at
$100\,\nicefrac{\mu C}{cm^{2}}$. The Neal temperature for BFO is
around 640 K ( 367 °C) and above this the structure tends towards
tetragonal \citep{karpinsky2017thermodynamic}. Beyond the ferroelectric
transition the structure resolves as cubic. Currently BFO and its
derivatives are some of the rare room temperature multi-ferroics and
as such have found many applications in nanoelectronic devices \citep{crassous2011nanoscale},
spintronic applications \citep{allibe2012room}, photovoltaics and
others \citep{hemme2021elastic}. However, single crystals are not
simple to grow and frequently it is more feasible to work with thin
films or ceramics \citep{wang2020structure}. Because of the volatility
of $Bi^{3+}$ and the aggregation of oxygen vacancies along poly crystalline
boundaries, it is not uncommon during sintering for the formation
of secondary phases, such as the mullite-like structured $Bi_{2}Fe_{4}O_{9}$
\citep{kirsch2019structural}. $Bi_{2}Fe_{4}O_{9}$ is an intriguing
multiferroic in its own right. Unlike $BFO$ is paramagnetic at room
temperature and it structure is very different, belonging to the orthorhombic
$Pbam$ space group \citep{kirsch2019structural}. Unlike $BFO$,
$Bi^{2+}$ ions sit in the channels formed by the $FeO_{6}$ octrahedra,
parallel to the c-axis. The result is that the stereochemically $6s^{2}$
Lone Pair Electrons (LEP), responsible for its rather poor ferroelectric
behaviour \citep{mackenzie2008electronic}, are similarly aligned
\citep{curti2013liebau}. Its Néel temperature is 264 K and at room
temperature it is antiferromagnetic. A comparison of unit cell sizes
of the two reveals that there is considerable mismatch between them
(see figure \ref{fig:A-comparison-of}). Furthermore, the net dipole
moment from the LEP is aligned along the c axis for the mullite-type
structure, rather than along the diagonal as for the perovskite structure
of $BFO$. One must assume that in mixed ceramics the domain boundary
between the two phases will be a source of rich dielectric behaviors. 

The dielectric permittivity of polycrystalline \citep{mackenzie2008electronic}
BFO/BFO derivatives and ceramics \citep{markiewicz2011dielectric}
and has been studied in the past with an emphasis on the high dielectric
permittivities that can be obtained with such materials .
In the case of Markiewicz et al. \citep{markiewicz2011dielectric}
the secondary phase was $Bi_{25}FeO_{40}$ and they noticed a rich
dielectric landscape. Most of the behaviour could be related to motion
along grain boundaries, along with distinct dielectric processes arising
from carrier hopping between $Fe^{2+}$ and $Fe^{3+}$ sites, as was
shown by X-ray Photoelectron spectroscopy or the short range hopping
of oxygen vacancies. In this study we concentrate on $BFO/Bi_{2}Fe_{4}O_{9}$
ceramics and consider the characteristics of the observed behaviours
in terms of the frequency and temperature trends, rather than in terms
of their dielectric constants, as is traditionally done \citep{mackenzie2008electronic},
\citep{perejon2019electrical}.

\section{Materials Preparation}

BFO synthesis results in a composition of three stable phases, depending
on the ratio between the Fe and Bi based molecules in the starting
materials, the synthesis temperature, the annealing history and impurities
\citep{orr2020sintering}. The three stable phases are $BiFeO_{4}$,
the mullite-like $Bi_{2}Fe_{4}O_{9}$ and sillenite-like $Bi_{25}FeO_{40}$.
In this work we are interested in the BFO and mullite-like phases
which are non-symmetric while avoiding the cubic $Bi_{25}FeO_{40}$.
In order to compare similar previous work in the field \citep{markiewicz2011dielectric}
we opted for a low temperature sol-gel based synthesis \citep{chakraborty2018highly}
that maintains the grain size in the tenths of nanometers.

$Bi(NO3)_{3}$ $5H_{2}O$ and $Fe(NO_{3})_{3}$ $9H_{2}O$ were dissolved
in distilled water forming a 0.015 mole solution of each. This was
followed by mixing both solutions and adding 0.3 cc of 70\% nitric
acid ($HNO_{3}$). The solution was stirred for half an hour after
which a 0.05 mole solution of tartaric acid was added followed by
another half an hour of stirring. The material was poured into a specially
assembled temperature controlled flask with a stirring mechanism where
the temperature was maintained at 100 °C for approximately 7 hours.
The resulting gel was dried at 100 °C until it became an orange brown
powder. Disks with 12.7 mm diameter were from 600 mg of powder using
a pressure of approximately 100MPa and then annealed at 650 °C for
2 hours. The annealing temperature was chosen, based on Selbach\textquoteright s
work \citep{orr2020sintering}, \citep{selbach2009thermodynamic}
which has demonstrated that, based on Gibbs energy, within the 447\textasciitilde 767
°C temperature range, $BiFeO_{3}$ is a metastable compound with $Bi_{2}Fe_{4}O_{9}$
and $Bi_{25}FeO_{40}$ being slightly more stable. Hence the material,
slowly transforms into those compounds. Due to the 1:1 ratio between
the Bi and Fe the mullite-like structure was expected to prevail.

\section{Methodology}

\subsection{XRD}

XRD phase structure acquisition of the synthesized material disks
prior to sintering and after, was carried out using a PANalytical
X'Pert Pro fitted with a Cu K\textgreek{a} source (\textgreek{l}=1.5406Å).
Phase analysis was based on Rietveld refinement routines (FullProf
Suite) \citep{rodriguez1990fullprof} with the peaks compared to those
generated using the Crystallography Open Database (COD) \citep{gravzulis2009crystallography}.
Particle size was estimated using parameters obtained by fitting of
the peaks in Matlab using the Scherrer equation\citep{para2021challenges}.

\subsection{Dielectric Measurements}

Silver electrodes of 4 mm diameter were deposited by vapour deposition
on the ceramic tablet of 0.48 mm thickness. Dielectric measurements
were made use a Novocontrol Broadband Dielectric Spectrometer based
on an Alpha Impedance Analyser \citep{AlphaHRdielectricAnalyserNovocontrol}
in the frequency range 0.2 Hz to 105 Hz. Temperature control was provided
a Quattro PID temperature controller (Novocontrol GmbH, Hamburg, Germany).
The initial temperature protocol was $20\text{°}C$ to $-120\text{°}C$
and then $-120\text{°}C$ to $200\text{°}C$. A further measurement
was made at 90°C under a bias voltage 0 V to -40 V, -40 V to 40 V
and 40 V to 0 V with a 4 V voltage step . The sample was cooled and
stored at room temperature for a month and the measurement repeated
in the same frequency range, but temperature range from $0\text{°}C$
to $230\text{°}C$.

\section{Results and Discussion}

\subsection{XRD}

\begin{figure}[H]
\begin{centering}
\includegraphics[scale=0.7]{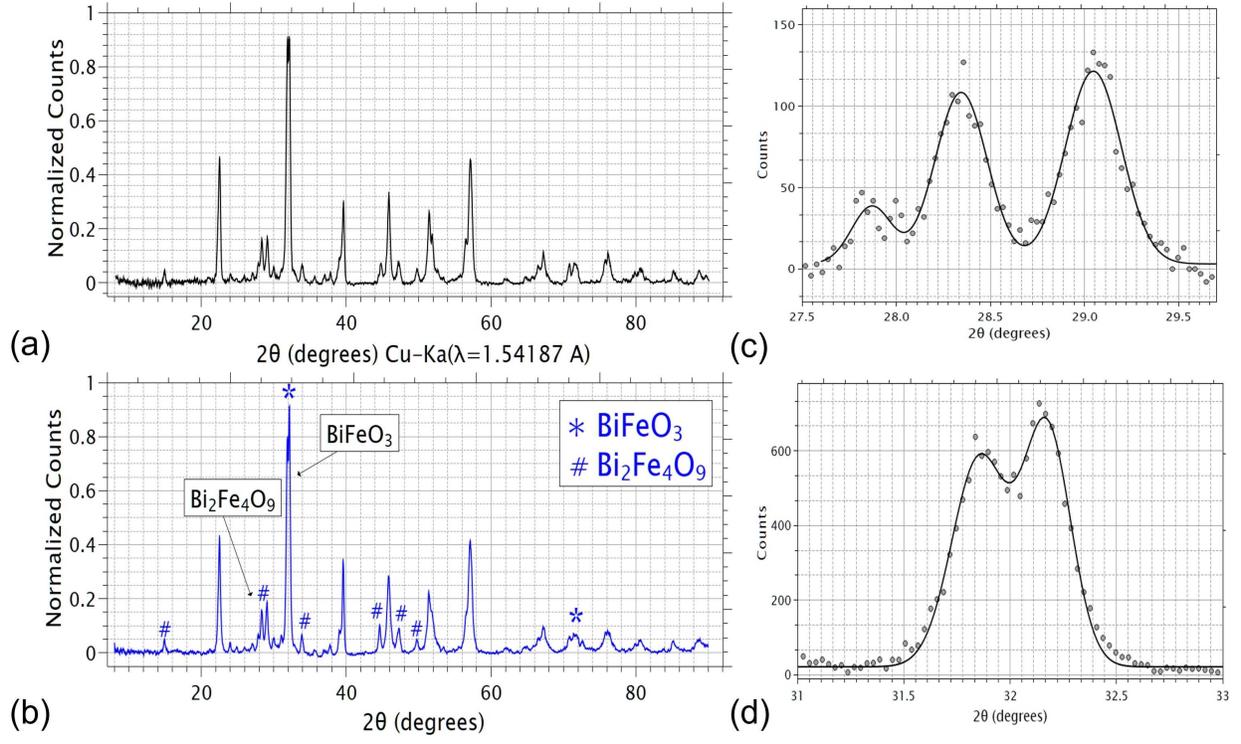}
\par\end{centering}
\caption{XRD patterns for a $BFO$ tablet. (a) powder, (b) sintered tablet.
The characteristic peaks for $BFO$ and $Bi_{2}Fe_{4}O_{9}$ are labelled,
(c) a gaussian fit for the $Bi2Fe_{4}O_{9}$ feature and (d) gaussian
fit for the BFO feature.}
\end{figure}

The XRD patterns for the powder and the sintered tablet are presented
in figure 3. A direct comparison of the patterns in figure (a) and
(b) reveals little change indicating the sintering did not change
composition. From Rietveld refinement \citep{para2021challenges}
the composition of the tablet is 70.5 \% $BFO$, 27.7\% $Bi_{2}Fe_{4}O_{9}$
and 1.8 \% $Bi_{25}FeO_{40}$. To assess particle size, the dominant
features of the two main constituents were fitted using gaussians
(shown in panels (c) and (d)).
\begin{equation}
f(\theta)=\sqrt{\frac{2}{\pi}}\cdot\frac{A_{i}}{\sigma_{i}}\cdot\exp\left(-2\left(\frac{2\theta-2\theta_{i}}{\sigma_{i}}\right)^{2}\right)\label{eq:f(theta)}
\end{equation}

Where Ai is the amplitude, i is the position and i is the width of
peak i. The results are listed in table \ref{tab:The-gaussian-fitting}.
Particle size is derived using the Scherrer equation \citep{patterson1939scherrer}
,
\begin{equation}
d=\frac{K\cdot\lambda}{\sigma_{i}\cdot\cos(\theta_{i})}\label{eq:Particle_size}
\end{equation}

Where $K$ is the Scherrer shape factor, usually taken as 0.92 for
cubic grains, and is the X-ray wavelength (1.54187 Å). The size range
calculated by equation 2 is shown in table 1. BFO crystallites are
notably larger than the secondary phase $Bi_{2}Fe_{4}O_{9}$.

\begin{table}[H]
\begin{centering}
\begin{tabular}{cccccccc}
\hline 
 & $A_{1}$ & $A_{2}$ & $2\theta_{1}$ & $2\theta_{2}$ & $\sigma_{1}$ & $\sigma_{2}$ & Size (nm)\tabularnewline
\hline 
$BFO$ & 179.7 & 180.2 & 31.8 & 32.17 & 0.257 & 0.225 & 31.5-36.9\tabularnewline
$Bi_{2}Fe_{4}O_{9}$ & 36.5 & 44.2 & 28.35 & 29.05 & 0.276 & 0.298 & 27-29.7\tabularnewline
\hline 
\end{tabular}
\par\end{centering}
\caption{\label{tab:The-gaussian-fitting}The gaussian fitting parameters of
figure 3(c) and (d)}
\end{table}

\subsection{Dielectric Permittivity}

\begin{figure}[H]
\begin{centering}
\includegraphics[scale=0.7]{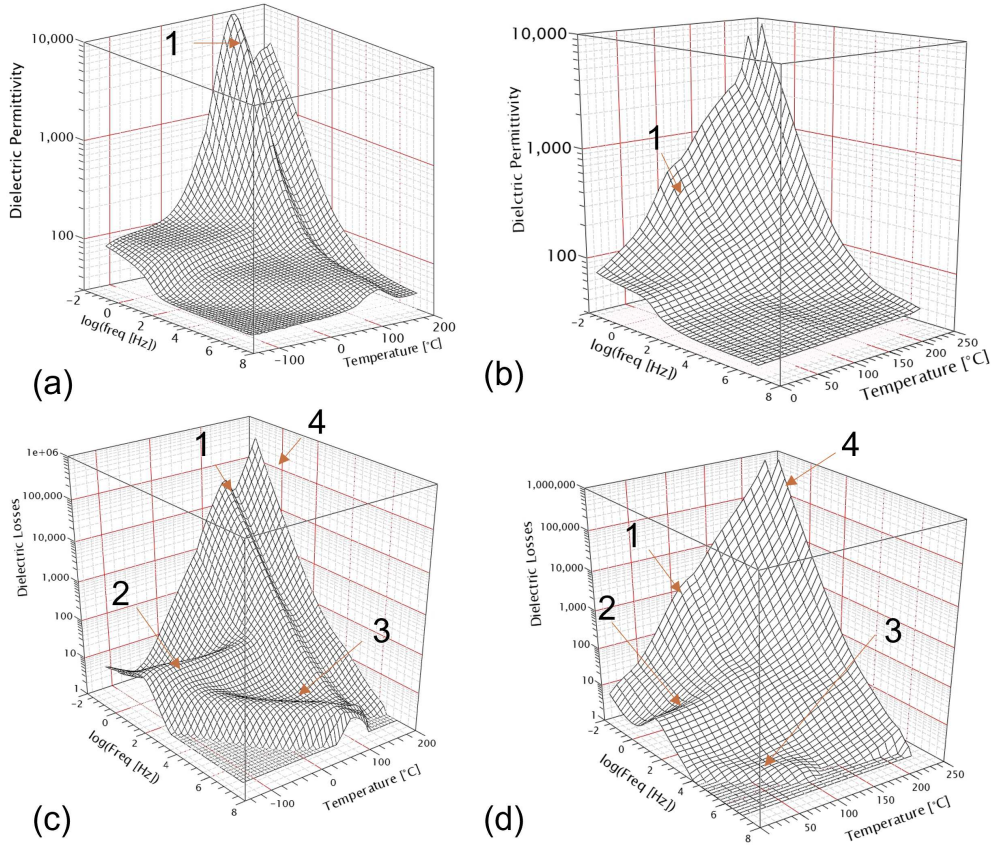}
\par\end{centering}
\caption{\label{fig:The-dielectric-permittivity}The dielectric permittivity
for the first measurement (a) and for the second measurement (b) of
the BFO tablet. In the second row are the corresponding dielectric
losses, (c) and (d) respectively. The highlighted features of the
relaxation landscape are: a second order phase transition (1), an
Arrhenius relaxation process with low activation energy (2), a Vogel-Fulcher-Tammenn
type relaxation process (3), dc conductivity overlaying low frequency
relaxations outside the measurement window (4).}
\end{figure}

The dielectric permittivities and losses for the BFO ceramic tablet
are presented in Figure \ref{fig:The-dielectric-permittivity}. The
upper graphs show the dielectric permittivity (a) for the initial
measurement and for the final measurement (b). The dielectric losses
are shown in the bottom graphs respectively.

It is apparent that the dielectric landscape is rich, with two relaxations,
marked 2 and 3 in the figure, inside the measurement window and dc
conductivity is present in the higher temperatures. One notes a phase
transition, marked 1 in the figure, with $T_{c}=100.6{^\circ}C$ (373
K). The transition temperature was established by finding the extremum
in the temperature derivative of the dielectric permittivity. While
the feature is still present in the second measurement it is greatly
diminished, indicating that continued reheating of the tablet has
led to a further modification of the phases. The bias measurement
around 90 °C, close to the phase transition temperature of 100.6 °C,
revealed no hysteresis. The dielectric permittivity at 0.13309 Hz
is presented in Figure 5(a). The vicinity of the phase transition
was modelled using the Curie-Weiss expression , $\varepsilon'(T)=C_{i}/(T-T_{C})$
, where $i=1,2$ for temperatures below and above $T_{c}$ respectively
\citep{lines2001principles}, \citep{strukov2012ferroelectric}. According
to Landau\textquoteright s phenomenological theory for phase transitions
\citep{lines2001principles} the ratio, $C_{1}/C_{2}$ , depends on
whether the phase transition is first order ($C_{1}/C_{2}=-8$) or
second order ($C_{1}/C_{2}=-2$ ) In our case it is -1.64, close to
-2 indicating that this is a second order structural ferroelectric
phase transition, consistent with the lack of hysteresis noted in
the bias measurement. MacKenzie et al. \citep{mackenzie2008electronic}
also noted an anomaly in the dielectric permittivity of $Bi_{2}Fe_{4}O_{9}$
at 252 °C, which they associated with the reported magnetic transition
temperature in pure $Bi_{2}Fe_{4}O_{9}$. However, this is far from
the anomaly reported here. Another possibility is that it is related
to the BFO phase. While the main ferroelectric phase transition is
at 825 °C \citep{catalan2009physics}, GHz spectroscopy of nearly
pure BFO also found phase transition at 130 °C \citep{krainik1966phase}.
Catalan et al. suggested that this was due to the presence of secondary
phases \citep{catalan2009physics}. Likewise, the mismatch between
$BFO/Bi_{2}Fe_{4}O_{9}$ crystallines can lead to large strains, as
has been noted in other crystalline systems \citep{carpenter1998spontaneous}.
Figure 6 of ref \citep{catalan2009physics} shows that a pressure
of 4 GPa would be enough to reduce $T_{c}$ in $BFO$ to 100 °C and
this is easily achievable on the local level of the domain boundary
\citep{carpenter1998spontaneous}. Therefore, we speculate that this
transition is a locally induced strain effect on $BFO$ crystallites
in the immediate vicinity of $Bi_{2}Fe_{4}O_{9}$ crystallites.
\begin{figure}[H]
\begin{centering}
\includegraphics[scale=0.7]{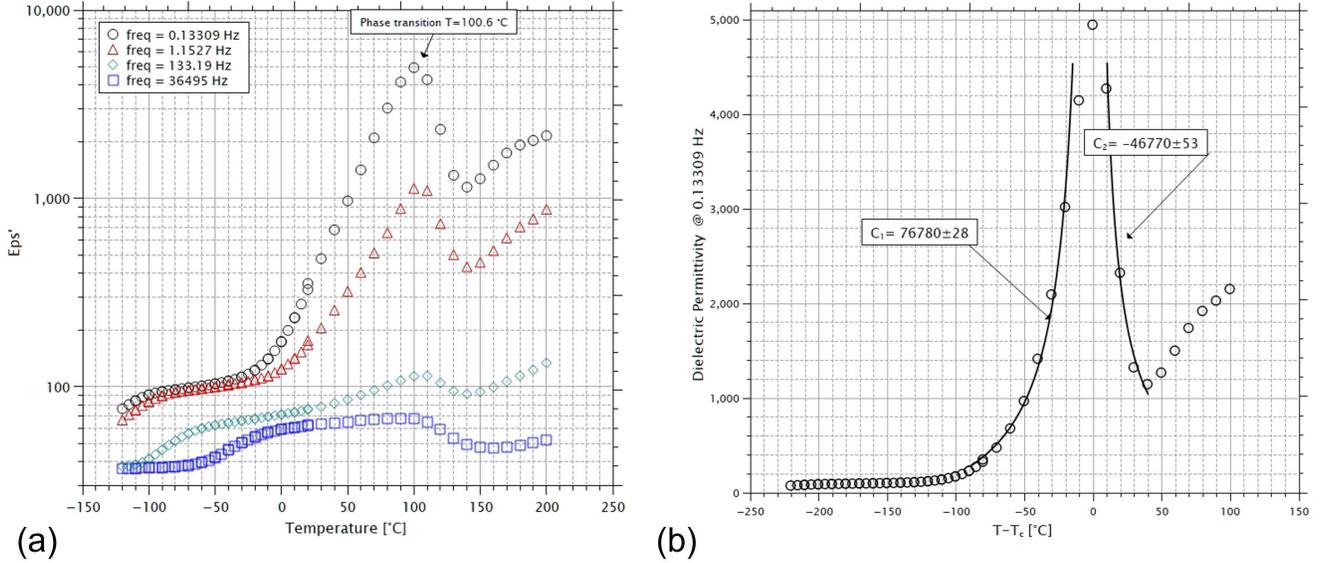}
\par\end{centering}
\caption{(a) The dielectric permittivity as a function of temperature for 4
frequencies, showing a dominant phase transition at $T_{c}=100.6{^\circ}C$
and (b) Curie-Wiess fitting before and after the phase transition
temperature, $\varepsilon'(T)=C_{i}/(T-T_{c})$. The ratio of the
constants is $C_{1}/C_{2}--1.64$, indicating a second order ferroelectric
phase transition.}
\end{figure}

The complex dielectric permittivity, $\varepsilon^{*}(i\omega)=\varepsilon'(\omega)-i\varepsilon(\omega)$,
where $\varepsilon'(\omega)$ is the dielectric permittivity and $\varepsilon(\omega)$
is the dielectric losses, was modeled using a sum of 3 Havrilak-Negami
(HN) functions along with a dc conductivity term,
\begin{equation}
\varepsilon^{*}(i\omega)=\varepsilon_{\infty}+\Sigma_{j=1}^{3}\frac{\Delta\varepsilon_{j}}{\left(1+(i\omega\tau_{j})^{\alpha_{j}}\right)^{\beta_{j}}}+\frac{\sigma_{dc}}{i\omega\varepsilon_{0}}\label{eq:Complex_permittivity}
\end{equation}

Where $\varepsilon_{\infty}$ is the higher frequency contribution
to the dielectric permittivity, $\Delta\varepsilon_{j}$ is the dielectric
strength of the $j$th process, $\tau_{j}$ is the corresponding characteristic
relaxation time, $0<\alpha_{j},\,\beta_{j}\leq1$ are the shape parameters,
$\sigma_{dc}$ is the dc conductivity, $\omega$ is the cyclic frequency,
$i$ is the root of -1 and $\varepsilon_{0}=8.85\times10^{-12}\,F/m$
is the permittivity of free space. In the first set of measurements
the $3^{rd}$ process was used to model the tail of relaxations situated
at frequencies lower than the bounds of the measurement window but
intruding to the data. For the second measurement at high temperatures
(above 0 °C) 3 HN processes are found inside the measurement window
and so lower frequency processes are modelled using a Left-hand Jonscher
function \citep{jonscher1996universal}, $A\cdot(i\omega)^{-n}$,
where $A$ is the amplitude and $0<n<1$. The data was modelled using
an in house fitting routine, Datama \citep{axelrod2004dielectric},
based on Matlab \citep{mathworks2000matlab}. The software is capable
of fitting in the complex plain and exploits a logarithmic measure
for the merit function $\chi^{2}\left(\varepsilon^{*}(i\omega_{n}),\varepsilon^{f}\left(i\omega_{n},\{x_{j}\}\right)\right)=\Sigma_{n,j}\left[\log\left(\varepsilon^{*}(i\omega_{n})\right)-\log\left(\varepsilon^{f}\left(i\omega_{n},\{x_{j}\}\right)\right)\right]^{2}$,
where $\varepsilon^{*}(i\omega_{n})$ is the measurement data at the
frequency point $\omega_{n}$, $\{x_{j}\}$ is the parameter set defined
in equation \ref{eq:Complex_permittivity} and $\varepsilon^{f}\left(i\omega_{n},\{x_{j}\}\right)$
is the model function value for the parameter set and respective frequency
point. The advantage is that for data with values varying by orders
of magnitude over the frequency range, equal weight is given in the
merit function for all frequency points, leading to a more accurate
fit. The dc conductivity for both measurement runs is presented in
Figure \ref{fig:An-Arrhenius-plot}.

\begin{figure}[H]
\begin{centering}
\includegraphics[scale=0.7]{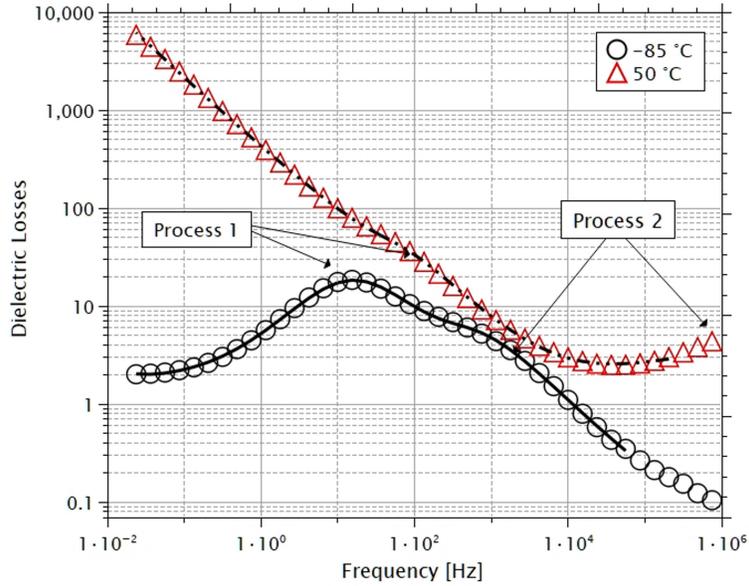}
\par\end{centering}
\caption{\label{fig:The-dielectric-losses}The dielectric losses at -85 °C
(circles) and at 50 °C (triangles) for the first measurement run.
The solid lines are the fitting function described by equation \ref{eq:Complex_permittivity}.
The 2 dominant processes are labeled}
\end{figure}
\begin{figure}[H]
\begin{centering}
\includegraphics[scale=0.7]{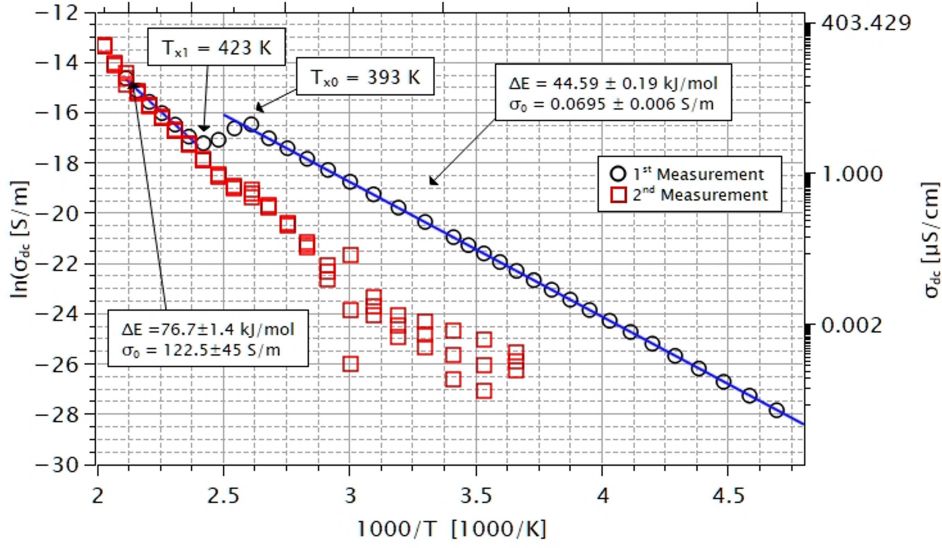}
\par\end{centering}
\caption{\label{fig:An-Arrhenius-plot}An Arrhenius plot of the dc conductivity
for the first (black circles) and second (blue squares) measurement
runs. The fitting lines are according to the Arrhenius expression.
The sample undergoes a compositional change during the first run between
T=393 K and T=423 K. }
\end{figure}

In both cases the conductivity is an activated process and obeys the
Arrhenius expression for conductivity,
\begin{equation}
\sigma_{dc}(T)=\sigma_{0}\exp\left(\frac{-\Delta E}{k_{B}T}\right)\label{eq:ArrheniusExpressionOfConductivity}
\end{equation}

where $\sigma_{0}$ is the high temperature limit of conductivity,
$T$ is the temperature in Kelvin, $k_{B}$ is Boltzmann\textquoteright s
coefficient and is the energy of activation for the process. The parameters
are listed in Table \ref{tab:The-parameters-of-the-Arrhenius-fit-for-dc-conductivity}.
At T=393 K the sample undergoes a compositional change, stabilizing
at T=423 and resulting in a doubling of the energy of activation.
The dc conductivity during the second measurement run matches these
new parameters. Usually dc conductivity in perovskite systems. Like
$BFO$ are assigned to the migration of oxygen vacancies \citep{markiewicz2011dielectric},
\citep{perejon2019electrical}. In mullite-like $Bi_{2}Fe_{4}O_{9}$
it has been assigned to intrinsic holes, centered around Oxygen. Reducing
$Bi_{2}Fe_{4}O_{9}$ by oxygen loss led to an extrinsic n-type semiconductor
via $O^{2-}\rightarrow\frac{1}{2}O_{2}+2e^{-}$ \citep{perejon2019electrical}.
However, In the case of $BFO$ the energies of activation are usually
around 0.9 eV to 1.1 eV (87 kJ/mol to 106 kJ/mol), considerably higher
than those measured here, although for spark plasma sintering (SPS)
of $BFO$ a value of $0.67\pm0.07eV$ ($64.6\pm6.8\,kJ/mol$) was
found \citep{perejon2013electrical}. In the sillenite like phase
$Bi_{25}FeO_{40}$ dc conductivity has been measured with an activation
energy of $0.52\pm0.02eV$ ($50.2\pm1.9\,kJ/mol$) \citep{perejon2019electrical}
. Given the very low percentage of this phase in the sample (1.8 \%),
it is unlikely that this is the source of conductivity we see. The
value of conductivity at 80 °C is $1.8\times10^{-7}\,S/m$ for the
first run and $5.8\times10^{-10}S/m$ for the second run. For both
runs dc conductivity was $4.4\times10^{-7}\,S/m$ at 200 °C. This
can be compared to the reported values of $2.77\times10^{-7}S/m$
for pure $Bi_{25}FeO_{40}$ ceramic \citep{perejon2019electrical}
at 350 °C and more than $2\times10^{-~6}~S/m$ for $BFO$ ceramic
\citep{perejon2013electrical}. The origin of the low activation energy
in SPS BFO was attributed to the reduction, caused by the sintering
process itself, along grain boundaries. The removal of oxygen from
the sample led to an excess of electrons, which in turn led to a reduction
of conducting holes by recombination. One can assume that in our sample
a similar process had occurred at the grain boundaries between the
two dominant compositions. This is further enhanced as a hypothesis
by the dramatic change in conductivity starting a 393 K in the first
run. The Novocontrol system maintains the sample temperature by a
temperature controlled Nitrogen flow, possibly leading to further
oxygen loss in the sample and activation energies close to those reported
to SPS BFO.

\begin{table}[H]
\begin{centering}
\begin{tabular}{ccc}
\hline 
 & Measurement 1 (T<393K) & Measurement 1\&2 (T>423K)\tabularnewline
\hline 
$\sigma_{0}\,[S/m]$ & $0.0695\pm0.0006$ & $123\pm45$\tabularnewline
$\Delta E\,[kJ/mol]$ & $44.59\pm0.19$ & $76.7\pm1.4$\tabularnewline
\hline 
\end{tabular}
\par\end{centering}
\caption{\label{tab:The-parameters-of-the-Arrhenius-fit-for-dc-conductivity}The
parameters of the Arrhenius fit for dc conductivity. }
\end{table}

\begin{figure}[H]
\begin{centering}
\includegraphics[scale=0.7]{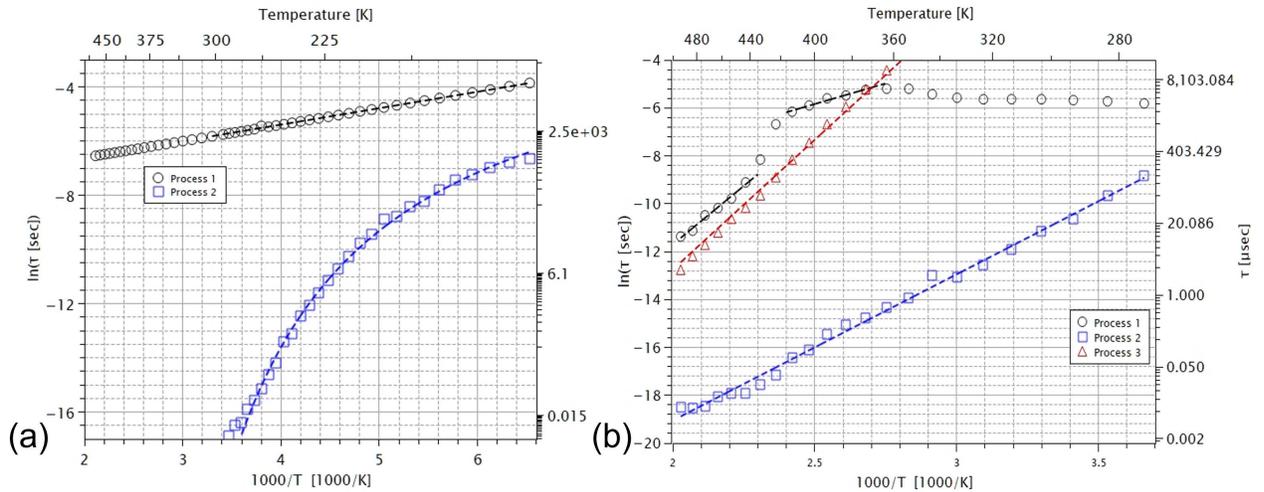}
\par\end{centering}
\caption{\label{fig:The-relaxation-times}The relaxation times in an Arrhenius
plot for the first measurement run (a) and for the second run (b).
The first run is dominated by two processes, one weakly temperature
activated with an Energy of Activation of $5.04\pm0.02\,kJ\cdot mol^{-1}$
. The second process follows VFT behaviour, indicative of a glassformer.
In the second measurement run Process 1 is still present, but greatly
modified. Above 360 K (approximately at the indicated Phase transition
temperature) it becomes temperature activated, undergoing a further
modification at 440 K. Additionally, there is no sign of the glassforming
process, but 2 strongly temperature activated processes appear. }
\end{figure}

The fitting reveals two distinct dielectric relaxations in the first
measurement and three distinct relaxations in the second, higher temperature
measurement. As stated above this indicates that the first temperature
run induced further sintering in the sample. The relaxation times
for the processes are presented in Figure \ref{fig:The-relaxation-times}.

The first temperature run, Figure \ref{fig:The-relaxation-times}(a),
is dominated by two processes (aptly marked Process 1 and Process
2 in the figure). Process 1 is Arrhenius with an energy of activation
of $\Delta E=5.04\pm0.02\,kJ\cdot mol^{-1}$. This range of activation
energy is typical of Maxwell- Wagner processes over domain boundaries
\citep{kremer2002broadband} and we can readily assign it to mismatched
grain boundaries in the ceramic. The second process is more puzzling.
At first glance it can be modeled by a Vogel Fulcher Tammann (VFT)
formula of the form
\begin{equation}
\tau(T)=\tau_{0}\exp\left(\frac{E_{0}}{k(T_{0}-T)}\right)\label{eq:tau_VFT}
\end{equation}

With the parameters $\tau_{0}=10^{14}\,s$, $E_{0}=199\,kJ\cdot mol^{-1}$
and $T_{0}=766\,K$. The VFT formula is usually associated with a
glass transition in dielectric science \citep{kremer2003scaling}.
However, in this case it can be linked to the dynamic rearrangement
of domain boundaries between different crystallites in the ceramic.
Thermodynamically the relaxation time is proportional to the probability
of a transition between two energy states to the system,
\begin{equation}
\tau^{-1}(T)\propto P\propto\exp\left(\frac{-\Delta F}{k_{B}T}\right)\label{eq:Probability_of_transition_between_two_states}
\end{equation}

Where $\Delta F$ is the Helmholtz free energy and $k_{B}$ is Boltzmann\textquoteright s
constant. Formally the Helmholtz energy is given by
\begin{equation}
F=\frac{1}{2}V\sum_{ij}\sigma_{ij}\varepsilon_{ij}-ST+\sum_{i}\mu_{i}N_{i}\label{eqn:HelmholtzEnergy}
\end{equation}
where, $V$ is the volume, $S$ is the entropy, $\sigma_{ij}$ and
$\varepsilon_{ij}$ are the stresses and strains between crystallites
$i$ and $j$ respectively, $\mu_{i}$ is the chemical potential of
the $i$ species and the $N_{i}$ is the number density. If the transitions
between crystallites of $BFO$ and $Bi_{2}Fe_{4}O_{9}$ are negligible,
then the source of the free energy is the mechanic energy caused by
the unit cell size mismatch at the grain boundaries (see Figure \ref{fig:A-comparison-of}).
Thermal energy in this case leads to rearrangements to reduce $F$
to its minimum,
\begin{equation}
\Delta F=\frac{1}{2}\left(\Delta V\underset{ij}{\sum}\sigma_{ij}\varepsilon_{ij}+V\underset{ij}{\sum}\Delta(\sigma_{ij}\varepsilon_{ij})\right)-T\Delta S+\underset{i}{\sum}\Delta(\mu_{i}N_{i})\label{eq:MinimumF}
\end{equation}

If the boundary strains are proportional to the unit cell sizes and
these do not vary, then the variation in free energy depends solely
on the changes in volume as the crystallites coalesce (the assumption
that there are negligible transition means that the last term in equation
\ref{eq:MinimumF} can be ignored). The volume $\Delta V$ is understood
to refer to any mesoscopic region where the relaxation takes place.
The probability of a transition then becomes
\begin{equation}
\tau^{-1}(T)\propto P\propto\exp\left(\frac{-dV\sum_{ij}\sigma_{ij}\varepsilon_{ij}-TdS}{k_{B}T}\right)=\exp\left(\frac{dS}{k_{B}}\right)\cdot\exp\left(-C\frac{dV}{k_{B}T}\right)\label{eq:ProbabilityOfTransitionVolume}
\end{equation}
One can assume that as the temperature is increased more internal
states are accessible, leading to an accelerated rate of change in
the respective internal crystalline volumes. An assumption of the
form $\Delta V\propto T/(T_{0}-T)$, leads to the modified VFT expression
of equation \ref{eq:tau_VFT}. The interpretation here is that as
the temperature $T_{0}$ is approached the mesoscopic volume tends
towards infinity, meaning that constriction effects imposed by the
boundary no longer play a significant role in the relaxation. Clearly,
as $T_{0}$ is a singularity in the expression the behaviour is not
valid in the immediate vicinity of $T_{0}$. $E_{0}$ is the energy
of the interface per unit surface area. 

This model is further reinforced by the relaxation times of the second
run, figure \ref{fig:The-relaxation-times}(b), where the VFT like
process 2 is missing. The second measurement run was only made from
0 °C to 230 °C. The $1^{st}$ process is still evident with approximately
the same time scales as the $1^{st}$ measurement run. It is largely
stationary until the phase transition noted at approximately 100 °C,
where it becomes Arrhenius. A further change is noted again at 170
°C (440 K) indicating a change in charge mobility along the grain
boundaries. A summary of the fitting parameters is presented in table
\ref{tab:The-fitting-parameters}. In addition, two more Arrhenius
processes are apparent in the fitting. The second process is replaced
a regular Arrhenius relaxation, and an additional relaxation appears
in the lower frequencies (noted as process 3 in figure \ref{fig:The-relaxation-times}(b)).
The energies of activation are in the range reported by Markiewicz
et al \citep{markiewicz2011dielectric} for hopping processes between
$Fe^{3+}$ and $Fe^{2+}$ traps induced by oxygen vacancies at grain
boundaries and it is most likely the same in our case.
\begin{table}[H]
\begin{centering}
\begin{tabular}{cccc}
\hline 
\textbf{Process} &  & \textbf{Activation Energy $[kJ/mol]$} & \textbf{Prefactor {[}s{]}}\tabularnewline
\hline 
Process 1 & $360K<T<420K$ & $28.7\pm2.4$ & $5.27\times10^{-7}$\tabularnewline
 & $440K<T<500K$  & $81.2\pm4.5$ & $2.71\times10^{-14}$\tabularnewline
Process 2 &  & $50.8\pm3.6$ & $2.64\times10^{-14}$\tabularnewline
Process 3 &  & $90.1\pm3.1$ & $1.14\times^{-15}$\tabularnewline
\hline 
\end{tabular}
\par\end{centering}
\caption{\label{tab:The-fitting-parameters}The fitting parameters for the
Arrhenius processes of Figure \ref{fig:The-relaxation-times}(b) }
\end{table}

\section{Conclusions}

$BFO$ and its derivatives hold great promise for applications. However,
they are not simple materials to produce because of the reactivity
of Bismuth. Understanding the dielectric picture can help decipher
an otherwise complicated mosaic. In this paper we have attempted to
do just that using dielectric spectroscopy and the language of dielectric
relaxation for a $BFO$ ceramic tablet consisting of 70.5 \% $BFO$,
27.7\% $Bi_{2}Fe_{4}O_{9}$ and 1.8 \% $Bi_{25}FeO_{40}$. The results
reveal an unstable sample where the sintering process is still active.
Grain boundary strains result if a partial onset of the Ferromagnetic
phase transition at temperatures far lower than expected. A kinetic
rearrangement of the sample is also revealed by the modification of
the dc conductivity of the sample starting from 370 K. Furthermore,
the non-Arrhenius relaxation in the first temperature run is explained
by way of a thermodynamic model related to the coalition of differing
sized crystalline particles. We feel that detailed dielectric studies
of these ceramics and their derivatives can lead to more robust sintering
regimes.

\bibliographystyle{elsarticle-num}
\bibliography{Complex_dielectric_behaviours_in_BFO_ceramics}

\end{document}